\documentclass[aps,prb,twocolumn,amsmath,amssymb,nofootinbib,superscriptaddress]{revtex4-1}

\usepackage{times}
\usepackage[pdftex]{graphicx}
\usepackage{dcolumn}
\usepackage{bm}
\usepackage{amsmath}
\usepackage{indentfirst}
\usepackage{float}
\usepackage[colorlinks]{hyperref}
\usepackage[dvipsnames]{xcolor}
\usepackage{xcolor}
\usepackage{subfigure}
\usepackage{algorithm}
\usepackage{algorithmicx}
\usepackage{algpseudocode}
\usepackage{amsmath}
\usepackage{verbatim}
\usepackage{makecell}
\usepackage[normalem]{ulem}

\newcommand{\Hop}{H}
\newcommand{\rhoeff}{\tilde{\rho}}
\newcommand{\sgx}{\sigma^x}
\newcommand{\sgy}{\sigma^y}
\newcommand{\sgz}{\sigma^z}
\newcommand{\sgp}{\sigma^+}
\newcommand{\sgm}{\sigma^-}

\newcommand{\trace}{{\rm tr}}

\newcommand{\rhoop}{\rho}
\newcommand{\rhoss}{\rho_{st}}
\newcommand{\lindblad}{\mathcal{L}}
\newcommand{\Leff}{\tilde{\mathcal{L}}}

\newcommand{\Heff}{\tilde{H}}
\newcommand{\im}{{\rm i}}
\newcommand{\dissipator}{\mathcal{D}}
\newcommand{\Dop}{C}
\newcommand{\Deff}{\tilde{C}}

\newcommand{\corr}{\mathcal{C}}
\newcommand{\Hc}{{\rm H. c.}}
\newcommand{\current}{\mathcal{J}}
\newcommand{\SVD}{{\rm SVD}}

\newcommand{\hnu}{Key Laboratory of Low-Dimensional Quantum Structures and Quantum Control of Ministry of Education, Department of Physics and Synergetic Innovation Center for Quantum Effects and Applications, Hunan Normal University, Changsha 410081, China
}

\begin{document}

\title{Density Matrix Renormalization Group Algorithm For Mixed Quantum States}

\author{Chu Guo}
\email{guochu604b@gmail.com}
\affiliation{Henan Key Laboratory of Quantum Information and Cryptography, Zhengzhou,
Henan 450000, China}
\affiliation{\hnu}



\begin{abstract}
Density Matrix Renormalization Group (DMRG) algorithm has been extremely successful for computing the ground states of one-dimensional quantum many-body systems. For problems concerned with mixed quantum states, however, it is less successful in that either such an algorithm does not exist yet or that it may return unphysical solutions. Here we propose a positive matrix product ansatz for mixed quantum states which preserves positivity by construction. More importantly, it allows to build a DMRG algorithm which, the same as the standard DMRG for ground states, iteratively reduces the global optimization problem to local ones of the same type, with the energy converging monotonically in principle. This algorithm is applied for computing both the equilibrium states and the non-equilibrium steady states, and its advantages are numerically demonstrated.
\end{abstract}

\maketitle

\section{Introduction}

Density Matrix Renormalization Group (DMRG) algorithm has become a standard numerical tool for computing the ground states of one-dimensional quantum many-body systems. As its defining feature, to compute the ground state of a many-body Hamiltonian, DMRG iteratively builds a local effective Hamiltonian for a single site (or two nearby sites) together with an effective environment compressed from the rest sites, and then computes the ground state of the local Hamiltonian. DMRG is well-known for its efficiency and extremely high precision in practice~\cite{White1992,White1993,Mcculloch2008,StauberHaegeman2019}. Compared to the imaginary time evolution algorithm, DMRG is free of errors resulting from time discretization as well as Trotter expansion~\cite{Trotter1959,Suzuki1976,PaeckelHubig2019}. During the last two decades, DMRG has been elegantly reformulated based on the variational Matrix Product State (MPS) ansatz for pure quantum state, and optionally the Matrix Product Operator (MPO) representation for the many-body Hamiltonian~\cite{Schollwock2011,Orus2014,ClaudiusSchollwock2017}. DMRG-like algorithms have also been applied to solve machine learning problems~\cite{StoudenmireSchwab2016,GuoPoletti2018,HanZhang2018,GuoPoletti2020,ShiGuo2022}.

For many problems of interest, the underlying quantum states are not pure states. In this work we will primarily consider two such instances: 1) the finite temperature equilibrium states (ESs) and 2) the non-equilibrium steady states (NESSs) of Lindblad equations. Similar to the MPS representation for pure states, mixed quantum states can generally be written as MPOs. The MPO representation for a quantum state is efficient if only a small \textit{bond dimension} is required for the MPO (the number of parameters for a generic matrix product ansatz usually grows quadratically with the bond dimension, and linearly with the system size). However a generic MPO does not guarantee positivity, which means that iterative algorithms built on variational MPO ansatz could easily result in unphysical solutions. This problem could be overcomed by the Matrix Product Density Operator (MPDO), which is a special form of MPO that is positive by construction~\cite{VerstraeteCirac2004}. In general MPO is more expressive than MPDO since if a density operator can be efficiently represented as an MPDO, then it can also be efficiently represented as an MPO, while the reverse may not be true~\cite{CuevasCirac2013}.

However, even if certain many-body mixed states can be efficiently represented as MPDOs~\cite{JarkovskyCirac2020}, there lacks a DMRG algorithm which directly works on a variational MPDO ansatz.  
For the NESS which is an eigenstate of a Lindblad operator $\lindblad$ with eigenvalue $0$, two approaches based on variational MPO ansatz have been used, which either substitute $\lindblad$~\cite{MascarenhasSavona2015,BaireyArad2020} or $\lindblad^{\dagger}\lindblad$~\cite{CuiBanuls2015,GuoPoletti2021,CasagrandeLandi2021} into the standard DMRG. Since $\lindblad$ is not Hermitian in general, convergence (even to local minima) is not guaranteed in the first approach.
In the second approach the nature of the original problem is completely ignored, the usage of $\lindblad^{\dagger}\lindblad$ will also introduce longer range interactions and square the bond dimension in the MPO representation. Moreover for problems with vanishing spectrum gaps for $\lindblad$~\cite{Znidaric2015}, the convergence could be extremely slow. 
To this end we note that a positive variational ansatz which takes into account nearest-neighbour correlaitons has also been applied to compute the NESSs~\cite{Weimer2015}, which is not in matrix product form and the algorithm is not DMRG-like.




The difficulty for building a DMRG algorithm directly on a variational MPDO ansatz is deeply related to the fact that the information about the mixedness (entropy) of a quantum state is \textit{global}, for example one can not tell whether an unknown quantum state is mixed or not by only performing local measurements on it. MPDO breaks the global mixed state into product of local mixed states. However, the ES, as an example, is formulated as a minimization problem that directly relies on the entropy. The local problem, if it can be formulated from a variational MDPO (or generally MPO) ansatz, is likely to lose the physical content that it originates from a globally mixed state and the nature of the local problem could be completely different from the global one. 


In this work we propose a variational positive matrix product ansatz (PMPA) for mixed states which overcomes the difficulty of MPDO. It can be seen as a very special form of MPDO with an orthogonal center at certain site $c$, similar to the mixed canonical representation of MPS. The orthogonal center is itself a proper mixed state on site $c$ and an effective environment. Most importantly, the orthogonal center is related to the global (mixed) quantum state by an isometry, thus it fully encodes the mixedness of the latter. 
We show that the standard DMRG algorithm can be straightforwardly generalized to work on the variational PMPA. The generalized DMRG algorithm is then applied for computing ESs and NESSs, and its advantages are numerically demonstrated against current state of the art algorithms, with either a much faster calculation or a higher precision.

\section{Variational positive matrix product ansatz} 
\begin{figure}
\includegraphics[width=\columnwidth]{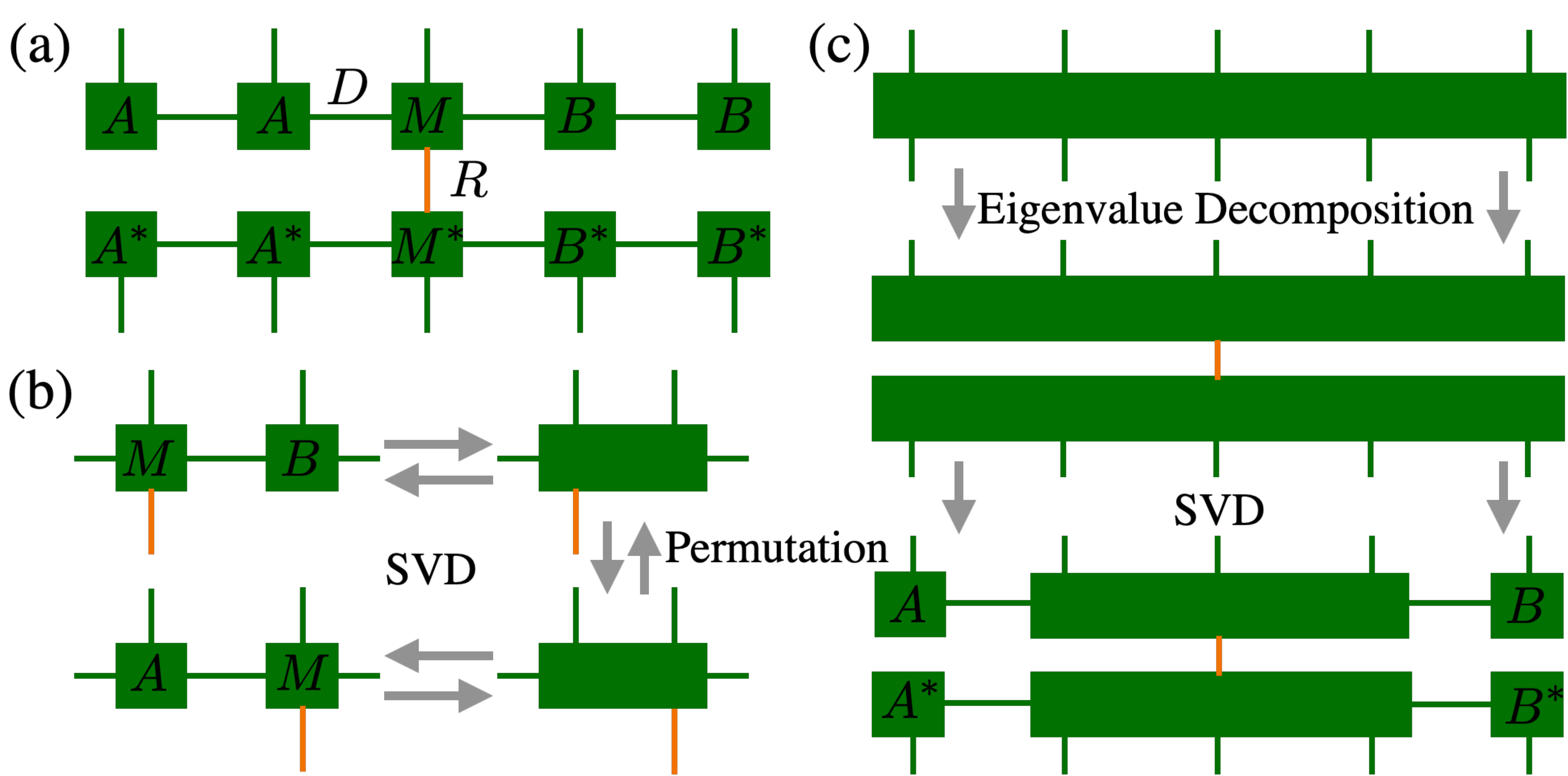}
\caption{(a) Positive matrix product ansatz for mixed quantum states. (b) Procedures to move the current orthogonal center to a nearby site: clockwise for the left-to-right sweep and anti-clockwise for the right-to-left sweep. (c) Preparation of a generic mixed state into the positive matrix product form.
}
\label{fig:fig1}
\end{figure}

The MPO representation of an $L$-site mixed quantum state is written as
\begin{align}\label{eq:mpo}
\rho_{s_1, \dots, s_L}^{s_1', \dots, s_L'} = \sum_{b_1, \dots, b_{L-1}} W^{s_1, s_1'}_{b_1} W^{s_2, s_2}_{b_1, b_2} \dots W^{s_L, s_L'}_{b_{L-1}},
\end{align}
with $s_l$ the physical index of size $d$, $b_l$ the auxiliary index and $W^{s_l, s_l'}_{b_{l-1}, b_l}$ the $l$-th site tensor. In the following we will omit the indices for $\rho$, and denote the $l$-th site tensor simply as $W_l$ when there is no confusion. 
The MPDO is a special form of MPO which guarantees positivity for each $W_l$ by requiring
\begin{align}\label{eq:mpdo_cond}
W^{s_l, s_l'}_{a_{l-1}, a_{l-1}', a_l, a_l'} = \sum_{\tau_l} M^{s_l, \tau_l}_{a_{l-1}, a_l} (M^{s_l', \tau_l}_{a_{l-1}', a_l'})^{\ast}.
\end{align}
Here the index tuple $(a_l, a_l')$ corresponds to $b_l$ in Eq.(\ref{eq:mpo}). The tensor $M_l$ in Eq.(\ref{eq:mpdo_cond}) can be interpreted as a site tensor (with two physical indices $s_l$ and $\tau_l$) of an MPS for a pure state (purification of $\rhoop$). 
The MPDO representation is certainly not unique. For example, one can fuse $\tau_l$ and $\tau_{l+1}$ into a single index $(\tau_l, \tau_{l+1})$ by contracting the two tensors $M_l$ and $M_{l+1}$, and then separating $(\tau_l, \tau_{l+1})$ into different $\tau_l'$ and $\tau_{l+1}'$ by splitting the resulting tensor using singular value decomposition (SVD). 



The PMPA is designed to ensure positivity and at the same time allows straightforward variational optimization. It can be seen as a very special form of MPDO which only allows a single $\tau_c$ at certain site $c$ (the `orthogonal center') to be non-trivial. Additionally, it requires that $W_l = A_l \otimes A_l^{\ast}$ for $l < c$ and $W_l = B_l \otimes B_l^{\ast}$ for $l > c$, where $A_l$ and $B_l$ are rank-$3$ tensors satisfying the left-canonical condition $\sum_{s_l, a_{l-1}} A^{s_l}_{a_{l-1}, a_l} (A^{s_l}_{a_{l-1}, a_l'})^{\ast} = \delta_{a_l, a_l'}$ and the right-canonical condition $\sum_{s_l, a_{l}} B^{s_l}_{a_{l-1}, a_l} (B^{s_l}_{a_{l-1}, a_l'})^{\ast} = \delta_{a_{l-1}, a_{l-1}'}$ respectively~\cite{Schollwock2011}. The PMPA is also shown in Fig.~\ref{fig:fig1}(a). 
The largest size of the auxiliary indices $a_l$ is referred to as the bond dimension of PMPA, denoted as $D = \max_{l}(\dim(a_l))$ similar to MPS. The size of the index $\tau_c$ is denoted as $R$, namely $R = \dim(\tau_c)$. The PMPA representation of $\rho$ is efficient if $D$ and $R$ remain almost unchanged when $L$ grows.  
Now defining the isometry
\begin{align}\label{eq:isometry}
V^{s_1, \dots, s_c, \dots, s_L }_{r_c, a_{c-1}, a_c} =&\delta_{s_c, r_c} \otimes  \sum_{a_1, \dots, a_{c-2}} A^{s_1}_{a_1} \dots A^{s_{c-1}}_{a_{c-2}, a_{c-1}} \otimes \nonumber \\ 
 & \sum_{a_{c+1}, \dots, a_{L-1}} B^{s_{c+1}}_{a_c, a_{c+1}}\dots B^{s_L}_{a_{L-1}}
\end{align}
and $\rhoeff_{s_c, a_{c-1}, a_c}^{s_c', a_{c-1}', a_c'} = W^{s_c, s_c'}_{a_{c-1}, a_{c-1}', a_c, a_c'} $ which simply reshuffles the indices of $W_c$, then the PMPA for $\rhoop$ can be written as
\begin{align}\label{eq:vpmpa}
\rho = V \rhoeff V^{\dagger}.
\end{align}
It follows that $\rhoeff$ has exactly the same spectrum property as $\rho$, therefore the mixedness of $\rho$ is fully encoded in $\rhoeff$. As such PMPA is only efficient for mixed states which are \textit{fairly pure}~\cite{GrossEisert2010}, that is, they can be written as the sum of a few pure states which can be efficiently represented as MPSs. More concretely, given a fixed integer $R$, the Schmidt rank of the underlying mixed state is bounded by $R$, and the entanglement entropy bounded by $\log(R)$. In other words, PMPA can not efficiently represent those mixed states whose entanglement entropy grows extensively with the system size, as a trivial example, a separable quantum state from the tensor product of local mixed states. Nevertheless, there also exists many interesting quantum states which are indeed fairly pure, to name a few examples, the low-temperature equilibrium states, the non-equilibrium steady states under certain engineered dissipation which drives the system towards a pure state~\cite{DiehlZoller2008,XuPoletti2018}, as well as the \textit{process tensor} that describes the multi-time quantum state of an open quantum system coupled to a finite environment~\cite{CostaShrapnel2016,PollockModi2018a,Guo2022a}. 

Given a (positive semidefinite) optimization problem on $\rhoop$, denoted as $f(\rhoop)$, a local problem $\tilde{f}$ of $\rhoeff$ naturally follows as
\begin{align}\label{eq:localf}
\tilde{f}(\rhoeff) = f(V \rhoeff V^{\dagger})
\end{align}
by keeping the isometry $V$ as constant. Importantly, due to the relation in Eq.(\ref{eq:vpmpa}), $\tilde{f}$ is often a same (positive semidefinite) optimization problem as $f$ as in the standard DMRG, which will be explicitly demonstrated in the applications later. After solving the local problem, one needs to move the orthogonal center as required by the DMRG sweep~\cite{Schollwock2011}, for which one can first contract the current center $M_c$ with the nearby site tensor $A_{c-1}$ or $B_{c+1}$ depending on the direction of the sweep, and then split the resulting two-site tensor (using SVD) with $\tau_c$ attached to the next center. This procedure is illustrated in Fig.~\ref{fig:fig1}(b). If the error induced by SVD is negligible, the new center will still be a proper mixed state and can be used as an initial guess for solving the next local problem. Therefore the local optimization can only improve the `energy' (value of $\tilde{f}(\rhoeff)$) since its solution should not be worse than the initial guess. With a well-defined local optimization problem and the center move technique, the standard DMRG algorithm for pure states can be straightforwardly generalized to mixed states. We will refer to this generalized DMRG algorithm as positive DMRG (p-DMRG) since it directly works on mixed states and preserves positivity. The initial PMPA for p-DMRG can be simply chosen as a randomly generated pure state in mixed canonical form (which is a PMPA with $R=1$). The p-DMRG algorithm is summarized in Algorithm.~\ref{alg:alg1}.

\begin{algorithm}[H]
\caption{Positive DMRG algorithm to find $\rhoop$ in positive matrix product form which minimizes $f(\rhoop)$.} \label{alg:alg1}
\begin{algorithmic}[1]
\State Initialize a random MPS with a fixed bond dimension $D$, and prepare it into right-canonical form\;
\For{$n = 1: \text{total number of sweeps}$}
\For{$c = 1:L-1$ } \Comment{left to right sweep}

\State Construct and solve the local optimization problem $\tilde{f}(\rhoeff_c)$ (Eq.(\ref{eq:localf})) \;
\State Compute $M_c$ by eigendecomposition of the optimal $\rhoeff_c$ and keeping only the $R$ largest Schmidt numbers (Eq.(\ref{eq:Mopt}));
\State Compute $\Psi_{a_{c-1}, a_{c+1}}^{s_c, \tau, s_{c+1}} = \sum_{a_c} M_{a_{c-1}, a_c}^{s_c, \tau} B_{a_c, a_{c+1}}^{s_{c+1}} $\;
\State Perform SVD on $\Psi$: $\SVD(\Psi_{a_{c-1}, a_{c+1}}^{s_c, \tau, s_{c+1}}) = \sum_{a_c} U_{a_{c-1}, a_c}^{s_c} \lambda_{a_c} V_{a_c, a_{c+1}}^{s_{c+1}, \tau} $ (Reserving $D$ largest Schmidt numbers) \;
\State Update $A_{a_{c-1}, a_c}^{s_c} \leftarrow U_{a_{c-1}, a_c}^{s_c} $\; 
\State Use $M_{a_c, a_{c+1}}^{s_{c+1}, \tau} = \sum_{a_c} \lambda_{a_c} V_{a_c, a_{c+1}}^{s_{c+1}, \tau}$ as the new orthogonal center on site $c+1$, which may be used as the initial guess for solving $\tilde{f}(\rhoeff_{c+1})$\;

\EndFor

\For{$c = L:2$}  \Comment{right to left sweep}


\State Construct and solve the local optimization problem $\tilde{f}(\rhoeff_c)$\;
\State Compute $M_c$ by eigendecomposition of the optimal $\rhoeff_c$ and keeping only the $R$ largest Schmidt numbers;
\State Compute $\Psi_{a_{c-2}, a_{c}}^{s_{c-1}, \tau, s_{c}} = \sum_{a_{c-1}} A_{a_{c-2}, a_{c-1}}^{s_{c-1}} M_{a_{c-1}, a_c}^{s_c, \tau}  $\;
\State Perform SVD on $\Psi$: $\SVD(\Psi_{a_{c-2}, a_{c}}^{s_{c-1}, \tau, s_{c}}) = \sum_{a_{c-1}} U_{a_{c-2}, a_{c-1}}^{s_{c-1}, \tau} \lambda_{a_{c-1}} V_{a_{c-1}, a_c}^{s_c}  $ (Reserving $D$ largest Schmidt numbers) \;
\State Update $B_{a_{c-1}, a_c}^{s_c} \leftarrow V_{a_{c-1}, a_c}^{s_c}$\; 
\State Use $M_{a_{c-2}, a_{c-1}}^{s_{c-1}, \tau} = \sum_{a_{c-1}} U_{a_{c-2}, a_{c-1}}^{s_{c-1}, \tau} \lambda_{a_{c-1}}$ as the new orthogonal center on site $c-1$, which may be used as the initial guess for solving $\tilde{f}(\rhoeff_{c-1})$\;

\EndFor

\EndFor

\Comment{Move the orthogonal center to the middle}
\For{$c = 1:L/2-1$} 
\State Construct and solve the local optimization problem $\tilde{f}(\rhoeff_c)$\;
\State Compute $M_c$ by eigendecomposition of the optimal $\rhoeff_c$ and keeping only the $R$ largest Schmidt numbers;
\State Compute $\Psi_{a_{c-1}, a_{c+1}}^{s_c, \tau, s_{c+1}} = \sum_{a_c} M_{a_{c-1}, a_c}^{s_c, \tau} B_{a_c, a_{c+1}}^{s_{c+1}} $\;
\State Perform SVD on $\Psi$: $\SVD(\Psi_{a_{c-1}, a_{c+1}}^{s_c, \tau, s_{c+1}}) = \sum_{a_c} U_{a_{c-1}, a_c}^{s_c} \lambda_{a_c} V_{a_c, a_{c+1}}^{s_{c+1}, \tau} $ (Reserving $D$ largest Schmidt numbers) \;
\State Update $A_{a_{c-1}, a_c}^{s_c} \leftarrow U_{a_{c-1}, a_c}^{s_c} $\; 
\State Use $M_{a_c, a_{c+1}}^{s_{c+1}, \tau} = \sum_{a_c} \lambda_{a_c} V_{a_c, a_{c+1}}^{s_{c+1}, \tau}$ as the new orthogonal center on site $c+1$, which may be used as the initial guess for solving $\tilde{f}(\rhoeff_{c+1})$\;

\EndFor

\State Construct and solve the local optimization problem $\tilde{f}(\rhoeff_{L/2})$\; \;
\State Compute $M_{L/2}$ by eigendecomposition of the optimal $\rhoeff_{L/2}$ \;
\State Return the PMPA formed by tensors $A_l$ with $1\leq l < L/2$, $M_{L/2}$ and $B_l$ with $L/2<l \leq L$\;

\end{algorithmic}
\end{algorithm}

We note that in the end we have used an additional half sweep from the left boundary to the middle site to avoid the \textit{boundary effect} which will be explained later. The p-DMRG algorithm optimizes a single site in each step, similar to the single-site DMRG algorithm. While after the single-site optimization, p-DMRG generates a two-site tensor, and then perform SVD on it, which is similar to the two-site DMRG algorithm. Due to these features, p-DMRG can also be directly used in presence of global quantum symmetries~\cite{SinghVidal2011,MccullochGulacsi2002,Weichselbaum2012}, since quantum number blocks could be adapted during the center move as in two-site DMRG.

We also note that a generic mixed state $\rhoop$ can be systematically prepared into a PMPA as shown in Fig.~\ref{fig:fig1}(c). That is, one first performs an eigenvalue decomposition on $\rho$ to get $\rho = U \Lambda U^{\dagger} $, then one performs a sequence of SVDs on $U$ to bring it into the desired matrix product form. In the next we explicitly demonstrate the p-DMRG algorithm for computing ESs and NESSs.

\section{Positive DMRG algorithm for equilibrium states}

The ES of temperature $T$ for a Hamiltonian $\Hop$ is the minimum of the free energy
\begin{align}\label{eq:fe_global}
\min_{\rhoop} \left(  \trace(\Hop\rhoop) - T S(\rhoop)  \right) 
\end{align}
with $S(\rhoop) = -\trace(\rhoop \log (\rhoop))$ the Von Neumann entropy~\cite{NielsenChuang}. The solution is $\rho_T \propto \exp(-\beta \Hop)$ up to a normalization factor $\trace(\exp(-\beta \Hop))$, with $\beta=1/(k_B T)$ the inverse temperature ($k_B$ is the Boltzmann constant). Substituting Eq.(\ref{eq:vpmpa}) into Eq.(\ref{eq:fe_global}), we obtain the local problem as
\begin{align}\label{eq:fe_local}
\min_{\rhoeff}\left(\trace(\Heff \rhoeff) - T S(\rhoeff)\right),
\end{align}
where $\Heff = V^{\dagger} H V $ is exactly the local effective Hamiltonian in the standard DMRG. The solution $\rhoeff$ of Eq.(\ref{eq:fe_local}) is simply
\begin{align}\label{eq:fe_local_solution}
\rhoeff \propto \exp(-\beta\Heff).
\end{align}
Then one can obtain the optimal $M_c$ by eigen decomposition of $\rhoeff$ and reserving the $R$ largest eigenvalues. The complexity of evaluating Eq.(\ref{eq:fe_local_solution}) is $O(d^2D^6)$ since $\Heff$ is of size $d D^2\times dD^2$. However for low temperature, we expect $R$ to be small. In this case $M_c$ can be computed much more efficiently as follows. First we compute the $R$ smallest eigenvalues of $\Heff$, that is $\Heff U = U \Lambda$, with $\Lambda$ a $R\times R$ diagonal matrix of these eigenvalues. Note that for this operation one does not have to explicitly build $\Heff$, but only needs to implement its operation on an input vector, as a common practice in DMRG. The complexity of this operation is only $O(dD^3)$. Then we have
\begin{align}\label{eq:Mopt}
M_c \propto U \exp(\frac{-\beta \Lambda}{2}).
\end{align}
Therefore the cost of solving each local problem is similar to that of the standard DMRG. In the zero temperature limit, $M_c$ becomes the ground state of $\Heff$.

To this end we note that in practice there are two effects that can hinder the monotonic convergence of the p-DMRG algorithm. First, there is a \textit{boundary effect} that is absent in the standard DMRG. Assuming $c=1$, then the size of $\Heff_1$ is at most $d^2 \times d^2$ since $\dim(a_1)\leq d$. If $d^2 < R$, then $\rhoeff_1$ does not have enough degrees of freedom to accommodate $R$ nonzero Schmidt numbers. As a result optimization at the boundaries could be less accurate and the energy may fluctuate. This effect can be avoided by grouping sites at the boundaries into larger ones~\cite{GarciaCirac2007}. Here we simply locate the final orthogonal center at the middle site to avoid this effect, which is done by the last half sweep in Algorithm.~\ref{alg:alg1}. Second, the SVD performed during the center move could also be a source of inaccuracy, similar to the SVD performed after a local optimization in the two-site DMRG algorithm. This effect could be leveraged by using a larger $D$.

\begin{figure}
\includegraphics[width=\columnwidth]{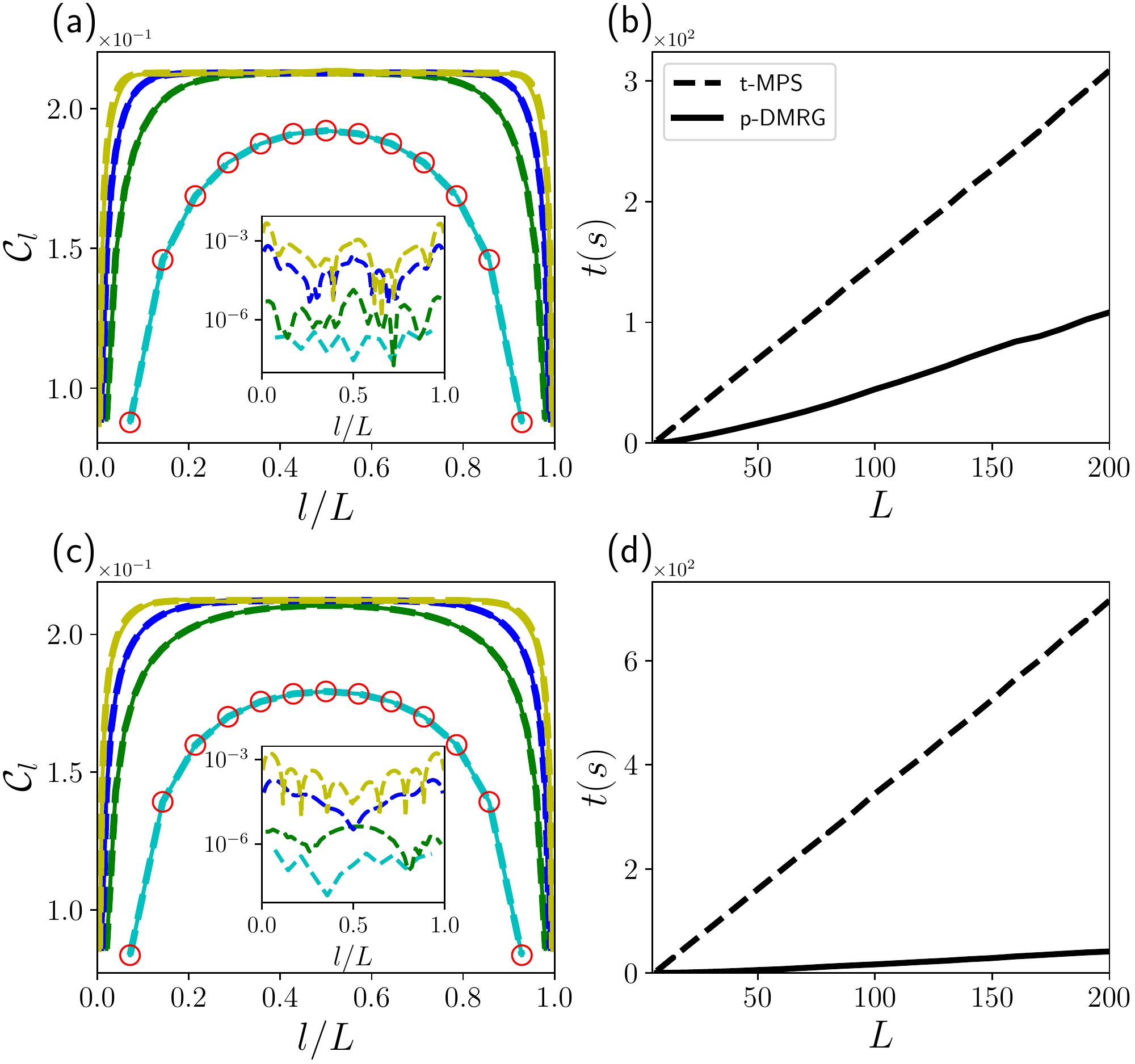}
\caption{(a,c) Correlation $\corr_l$ as a function of $l$ at $\beta=10$ (a) and $\beta=20$ (c). The cyan, green, blue, yellow solid (p-DMRG) and dashed (t-MPS) lines correspond to $L=14,50,100,200$ respectively. The insets show the average difference between p-DMRG and t-MPS results, defined as $\sum_{l=1}^{L-1}|\corr_l^{{\rm p-DMRG}} - \corr_l^{{\rm t-MPS}}| /(L-1)$. The red circles in (a,c) represent exact diagonalization results at $L=14$. (b,d) The runtime scaling for p-DMRG and t-MPS at $\beta=10$ (b) and $\beta=20$ (d). We have used $h=J=1$. For t-MPS we have used a $4$-th order Trotter expansion of $\exp(-\Hop_{Ising} d\tau)$ with step size $d\tau=0.05$. In both simulations we have used $D=30$. For p-DMRG we have further used $R=50$ for $\beta=10$ and $R=10$ for $\beta=20$, with $2$ sweeps for both cases. All the simulations done in this work use a single thread of a CPU with $3.5$ GHz frequency. 
}
\label{fig:fig2}
\end{figure}

To demonstrate the powerfulness of p-DMRG, we apply it to compute the low-temperature ES for the transverse field Ising chain and compare its performance with the imaginary time-evolving MPS (t-MPS) algorithm~\cite{Vidal2003,VerstraeteCirac2004,DaleyVidal2004,WhiteFeiguin2004}. The Hamiltonian is
\begin{align}\label{eq:ising}
H_{Ising} = h \sum_{l=1}^L \sgz_l + J \sum_{l=1}^{L-1} \sgx_l \sgx_{l+1},
\end{align}
with $\sgx$, $\sgy$, $\sgz$ the Pauli operators, $L$ the  total number of spins, $h$ the magnetization strength and $J$ the interaction strength (we set $h=1$ as the unit). The results are shown in Fig.~\ref{fig:fig2}, where we have computed the correlation
\begin{align}\label{eq:corr}
\corr_l = -\left(\trace(\sgp_l\sgm_{l+1}\rho_T) + \Hc\right)
\end{align}
at different sites (the correlation is chosen instead of an on-site observable such as $\trace(\sgz_l\rho_T)$ since it is in general much harder to converge). From Fig.~\ref{fig:fig2}(a, c) we can see that the p-DMRG results match very well with t-MPS results (difference of the order $10^{-3}$ at $L=200$). Meanwhile, p-DMRG has a more than $3$ and $21$ times speed up at $\beta=10$ and $\beta=20$ respectively, compared to t-MPS. Interestingly, while the t-MPS simulation becomes slower for lower temperature (since we need to evolve for longer times), the p-DMRG simulation becomes more efficient in the latter case.

\section{Positive DMRG algorithm for non-equilibrium steady states}

The NESS of a Lindblad equation, denoted as $\rhoss$, satisfies
\begin{align}\label{eq:lindblad}
\lindblad(\rhoss) = -\im [\Hop, \rhoss] + \sum_k \dissipator_k(\rhoss) = 0
\end{align}
with $\dissipator_k(\rhoop) = 2\Dop_k \rhoop \Dop_k^{\dagger} - \{\Dop_k^{\dagger}\Dop_k, \rhoop \} $~\cite{Lindblad1976,GoriniSudarshan1976}. $\rhoss$ is also a solution of the following minimization problem
\begin{align}\label{eq:lind_global}
\min_{\rhoop}\left| \frac{\trace(\rhoop^{\dagger} \lindblad\rhoop)}{\trace(\rhoop^{\dagger}\rhoop)} \right|.
\end{align}
Substituting Eq.(\ref{eq:vpmpa}) into Eq.(\ref{eq:lind_global}), we get the local problem
\begin{align}\label{eq:lind_local}
\min_{\rhoeff}\left| \frac{\trace(\rhoeff^{\dagger} \Leff\rhoeff)}{\trace(\rhoeff^{\dagger}\rhoeff)} \right|.
\end{align}
The explicit form of the local effective operator $\Leff$ is obtained by evaluating the numerator in Eq.(\ref{eq:lind_global}) term by term as
\begin{align}
&\trace(\rhoop^{\dagger} [\Hop, \rhoop]) = \trace(V\rhoeff^{\dagger}V^{\dagger}\Hop V\rhoeff V^{\dagger}) - \trace(V\rhoeff^{\dagger} V^{\dagger} V \rhoeff V^{\dagger}\Hop) \nonumber \\ 
=&\trace(\rhoeff^{\dagger} V^{\dagger} H V \rhoeff ) - \trace(\rhoeff^{\dagger}\rhoeff V^{\dagger}HV) = \trace(\rhoeff^{\dagger} [\Heff, \rhoeff]); \\
&\trace(\rhoop^{\dagger} \Dop_k \rhoop\Dop_k^{\dagger} ) = \trace(V\rhoeff^{\dagger}V^{\dagger}\Dop_k V\rhoeff V^{\dagger} \Dop_k^{\dagger}) = \trace(\rhoeff^{\dagger}\Deff_k \rhoeff \Deff_k^{\dagger}); \\
&\trace(\Dop_k^{\dagger}\Dop_k \rhoop) = \trace(\Dop_k^{\dagger}\Dop_k V \rhoeff V^{\dagger}) = \trace(V^{\dagger}\Dop_k^{\dagger}\Dop_k V \rhoeff ).
\end{align}
Here $\Heff=V^{\dagger}HV$ is the same to the local effective Hamiltonian for computing ESs, and $\Deff_k = V^{\dagger} \Dop_k V$. 
Combining all these terms together we thus obtain
\begin{align}\label{eq:Leff}
\Leff(\rhoeff) = -\im [\Heff, \rhoeff] + \sum_k(2\Deff_k \rhoeff \Deff_k^{\dagger} - \{V^{\dagger}\Dop_k^{\dagger}\Dop_k V, \rhoeff\}  ).
\end{align}
The second term in Eq.(\ref{eq:Leff}) is not in the standard Lindblad form (a standard Lindblad operator is the generator of some completely positive and trace preserving quantum map~\cite{SudarshanRau1961,JordanSudarshan1961}). Nevertheless, it has been shown that an operator in the form of Eq.(\ref{eq:Leff}) is a generator of a completely positive quantum map~\cite{Caves2000,LidarWhaley2001,NakazatoMessina2006}. The complexity of solving Eq.(\ref{eq:lind_local}) generally scales as $O(d^2D^6)$~\cite{LubaschBanuls2014b}. 

\begin{figure}
\includegraphics[width=\columnwidth]{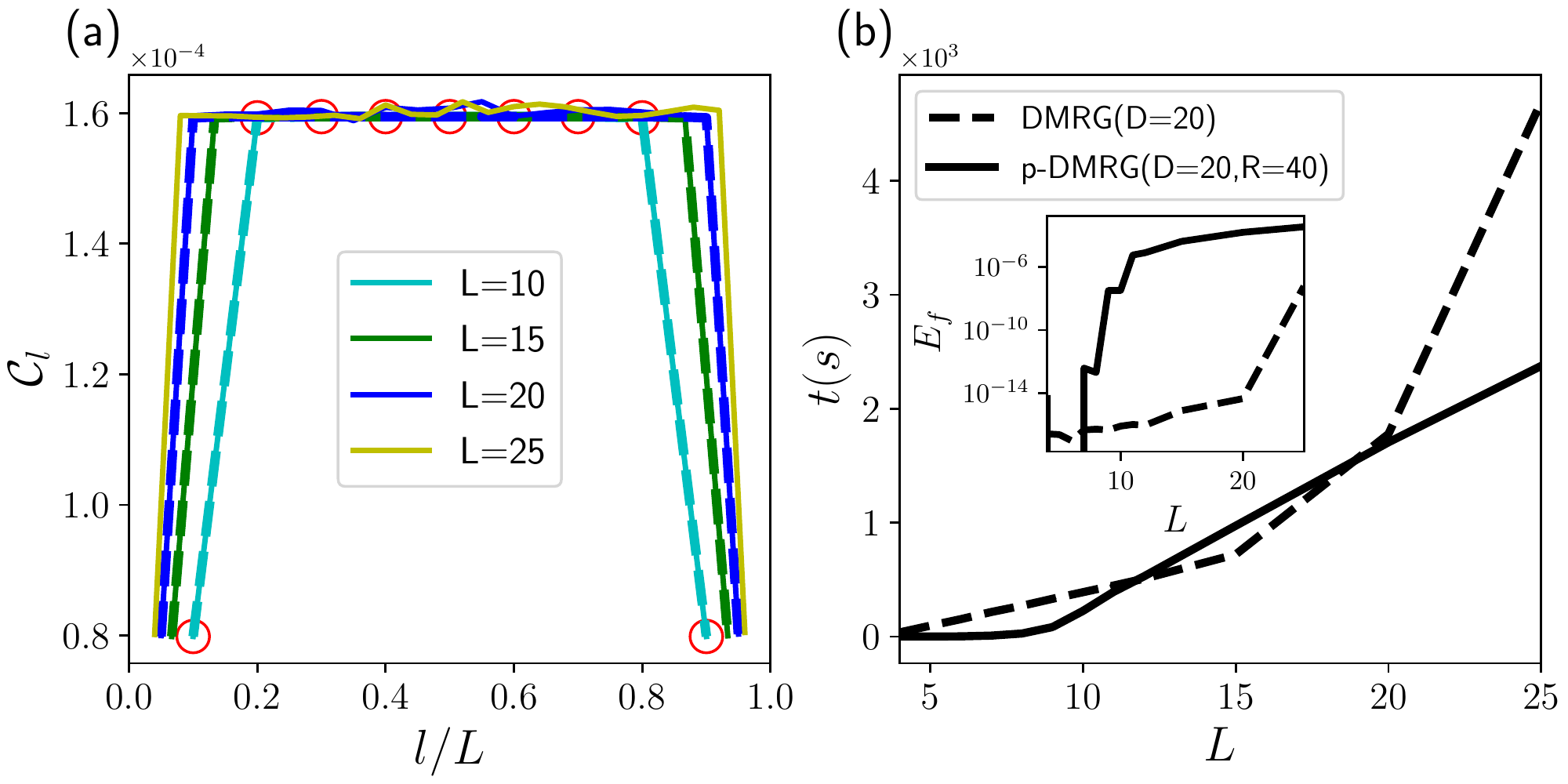}
\caption{(a) Correlation $\corr_l$ as a function of $l$. The cyan, green, blue, yellow solid lines represent p-DMRG results with $L=10,15,20,25$ respectively, while the cyan, green, blue dashed lines represent DMRG results with $L=10,15,20$ respectively. The red circles are results calculated with ED. (b) The runtime $t$ and the final energy $E_f$ (the inset) as a function of the system size $L$ for p-DMRG and DMRG respectively. $J=0.1$ and $h=1$ are used in our simulations. We have also used $10000$ sweeps for DMRG and $2$ sweeps for p-DMRG.
}
\label{fig:fig3}
\end{figure}

To demonstrate the p-DMRG algorithm for computing NESSs, we first study the dissipative Ising chain with the Hamiltonian in Eq.(\ref{eq:ising}) and the bulk dissipation~\cite{AtesLesanovsky2012,OverbeckWeimer2017}
\begin{align}\label{eq:ising_diss}
\dissipator_{Ising, l}(\rhoop) = 2\sgm_l \rhoop \sgp_l - \{\sgp_l\sgm_l, \rhoop\},
\end{align}
which acts on each spin and tends to drive it into the down state. Here we have used the anti-commutator $\{A, B\}=AB+BA$. The magnetization term (first term) in Eq.(\ref{eq:ising}) commutes with the dissipative term defined in Eq.(\ref{eq:ising_diss}), therefore for $J = 0$ the NESS of the dissipative Ising chain would simply be a separable state with each spin in the down state (thus a pure state). For $J \ll h = 1$, we expect that the underlying NESS would still be close to a pure state which could be suitable for our p-DMRG algorithm to solve. Similar to the equilibrium case, we compute the correlations defined in Eq.(\ref{eq:corr}) with DMRG (using $\lindblad^{\dagger}\lindblad$) and p-DMRG respectively, and the results are shown in Fig.~\ref{fig:fig3}. In Fig.~\ref{fig:fig3}(a), the DMRG results for $L=25$ are missing since they are clearly unphysical (for example, the local magnetization $\trace(\sgz_l \rhoss)$ on some sites become much smaller than $-1$).
In comparison the p-DMRG results are still in a reasonable range (although for $L=20$ the p-DMRG results are not as accurate as the DMRG results). This issue of DMRG still exists when using a larger bond dimension ($D=25$). Therefore this issue is more likely due to that DMRG has not fully converged for $L=25$ even after $10000$ sweeps (with the final energy $E_f$ already lower than $10^{-6}$ from Fig.~\ref{fig:fig3}(b)), instead of that the MPO ansatz with $D=20$ is not expressive enough (This issue of DMRG may be leveraged with a better initialization strategy~\cite{CasagrandeLandi2021}, while in this work we only consider random initialization for both DMRG and p-DMRG). For $J$ comparable to $h$, we find that DMRG can also obtain accurate solutions till $L=25$ (in comparison the p-DMRG results in such case using the same $D$ and $R$ become less inaccurate since the underlying NESS is more mixed).

As a second example, we demonstrate the p-DMRG algorithm for computing the NESS of a particularly hard (for numerical computation) problem: the boundary driven XXZ chain~\cite{Prosen2011a,Prosen2011b}, with the Hamiltonian
\begin{align}
\Hop_{XXZ} = \sum_{l=1}^{L-1}(\sgx_l\sgx_{l+1} + \sgy_l \sgy_{l+1} + \Delta \sgz_l\sgz_{l+1} ),
\end{align}
and the boundary dissipations
\begin{align}
\dissipator_{1}(\rhoop) &= 2\sgp_1 \rhoop \sgm_1 - \{\sgm_1\sgp_1, \rhoop\}; \\
\dissipator_L(\rhoop) &= 2\sgm_L \rhoop \sgp_L - \{\sgp_L\sgm_L, \rhoop\}.
\end{align}
Boundary driven open quantum systems provide important setups to study non-equilibrium transport problems~\cite{BertiniZnidaric2020,LandiSchaller2021}. In case the bulk system is integrable, the spectrum gap of $\lindblad$ typically scales as $1/L^3$~\cite{Znidaric2015,GuoPoletti2017a}, which makes it extremely difficult to compute the NESS even for small systems. Utilizing a special global U(1) symmetry of such systems~\cite{GuoPoletti2019}, exact diagonalization (ED) up to $14$ spins has been performed~\cite{GuoPoletti2017b}. Moreover, DMRG based on $\lindblad$ almost can never converge in this case (in comparison for bulk dissipative systems it can often quickly converge for even hundreds of spins~\cite{BaireyArad2020}), while DMRG based on $\lindblad^{\dagger}\lindblad$ converges extremely slowly and can easily be trapped~\cite{GuoPoletti2021}.

\begin{figure}
\includegraphics[width=\columnwidth]{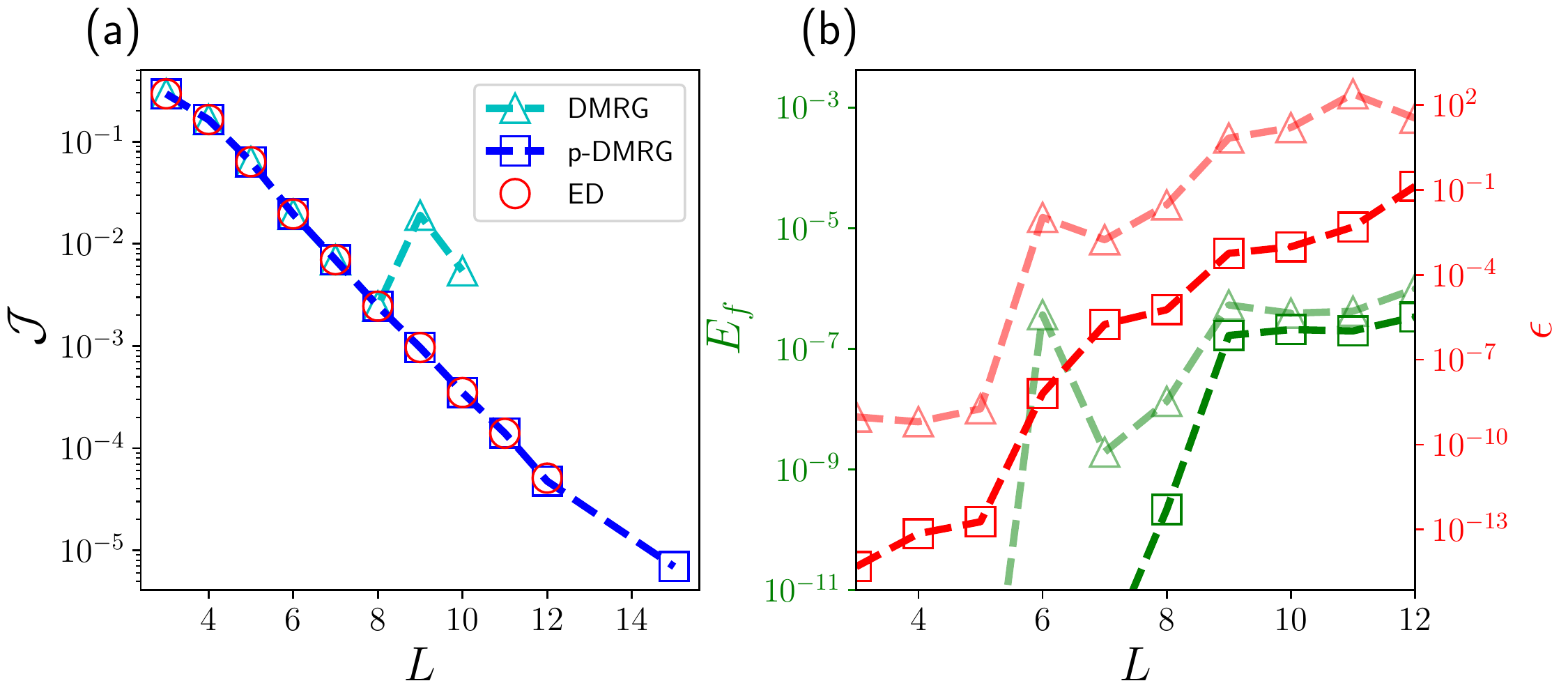}
\caption{(a) Scaling of the steady state current $\current$ with system size $L$, computed by DMRG, p-DMRG and ED. Since the results for (p-)DMRG may not converge, we use $\current_{L/2}$ ($\current_{(L-1)/2}$ if $L$ is odd) instead in these cases. (b) The green dashed lines corresponding to the left axis plot the final energy ($E_f$) for p-DMRG (square) and DMRG (triangle) as a function of $L$. The red dashed lines corresponding to the right axis plot the mean relative error of the current, defined as $\epsilon = \sum_{l=1}^{L-1} |\current_l/\current - 1|/(L-1)$ with $\current$ the exact value from ED, for p-DMRG (square) and DMRG (triangle) as a function of $L$. In all those simulations we have used $D=15$ and $10000$ sweeps for DMRG, $D=20$, $R=50$ and $2$ sweeps for p-DMRG. 
}
\label{fig:fig4}
\end{figure}

The numerical results are shown in Fig.~\ref{fig:fig4}, where we have computed the steady state current defined as
\begin{align}\label{eq:current}
\current_{l} = \im \left(\trace(\sgp_l\sgm_{l+1}\rhoss) - \Hc\right)
\end{align}
with ED, DMRG (using $\lindblad^{\dagger}\lindblad$) and p-DMRG respectively. $\current_l$ is $l$-independent if $\rhoss$ is exact. We focus on the strongly interacting scenario with $\Delta=1.5$, where $\current$ decreases exponentially with $L$ (insulating) and is the most numerically challenging regime~\cite{Prosen2011b} (The regimes with $\Delta \leq 1$ have been solved quite accurately for several tens of spins using DMRG~\cite{CasagrandeLandi2021}). From Fig.~\ref{fig:fig4}(a) we can see that p-DMRG results with $D=20$, $R=50$ and $2$ sweeps agree fairly well with exact results up to $L=12$ (still reasonable for $L=15$), while DMRG results with $D=15$ and $10000$ sweeps fails to converge as early as $L=9$. 
The major issue for DMRG is that the observables are very off even if the final energy is of the same order as p-DMRG (although the energies in these two cases have completely different meanings), and they do not seem to improve for larger $D$s or more sweeps. For ED $L=12$ is already the upper limit we can deal with using a personal computer (without using the global U(1) symmetry). More details of the simulations done here can be found in Appendix.~\ref{app:xxzmore}.


\section{Discussions}  
We have proposed a positive matrix product ansatz (PMPA) for mixed quantum states, and demonstrated a generalized DMRG algorithm (p-DMRG) which directly works on the variational PMPA and preserves positivity. The advantages of p-DMRG for both equilibrium states and non-equilibrium steady states are numerically demonstrated with comparisons to state of the art algorithms. 

To this end we stress again that PMPA is only efficient for quantum states that are fairly pure. It is less expressive than MPDO and MPO since it belongs to a very special case of the latter ones. Therefore it is complementary, instead of a substitution, to existing algorithms based on MPDO and MPO.
As a trivial example, PMPA fails to (efficiently) represent the maximally mixed state (for which it requires $R=d^L$), which however can be efficiently represented as an MPO with bond dimension $1$. Therefore if the solution is close to the maximally mixed state, t-MPS algorithm which directly starts from the maximally mixed state is the method of choice~\cite{ProsenZnidaric2009}. Nevertheless, PMPA can efficiently represent many physically relevant quantum states as pointed out in the main text and demonstrated in our numerical examples. 

PMPA could also be useful in other settings such as quantum information, as a convenient ansatz that allows to efficiently compute many important quantities. For example, given two quantum states $\rho$ and $\sigma$ that can be efficiently represented as PMPAs, the quantum fidelity, defined as $\mathcal{F}(\rho, \sigma) = \trace^2(\sqrt{\sqrt{\rho}\sigma\sqrt{\rho}})$, can be efficiently computed, which is not possible even if both $\rho$ and $\sigma$ can be efficiently written as MPOs or MPDOs. 
Interestingly, an unknown quantum state, if assumed to be efficiently representable as PMPA, allows efficient quantum tomography~\cite{CramerLiu2010,LanyonRoos2017}.
Generalization of PMPA to higher dimensions could be interesting but also more challenging, for example, a canonical form of the PMPA could not be easily defined as in the one-dimension case (thus Eq.(\ref{eq:vpmpa}) and the nice properties of the PMPA would no longer hold). 
 



The code for the p-DMRG algorithm together with the examples used in this work can be found at~\cite{PDMRG}. C. G. acknowledges support from National Natural Science Foundation of China under Grant No. 11805279.


\bibliographystyle{apsrev4-1}
\bibliography{refs}

\appendix

\section{Efficient construction of the local problem for computing the equilibrium states}
As a standard practice in DMRG algorithm, the local problem $\tilde{f}(\rhoeff)$ can be efficiently constructed by reusing a large portion of the previous calculations~\cite{Schollwock2011}. In case of equilibrium states, it reduces to the construction of $\Heff$ as shown in the main text, which is exactly the same as the standard DMRG. We will sketch the procedures to build $\Heff$ here for completeness.

Assuming that the $L$-site many-body Hamiltonian $\Hop$ can be written as an MPO
\begin{align}\label{eq:ham}
\Hop^{s_1', s_2', \dots, s_L'}_{s_1, s_2, \dots, s_L} = \sum_{b_1, b_2, \dots, b_{L-1} } O_{b_1}^{s_1, s_1'} O_{b_1, b_2}^{s_2, s_2'} \dots O_{b_{L-1}}^{s_L, s_L'},
\end{align}
with $s_l$ the physical index and $b_l$ the auxiliary index. The largest size of the auxiliary index is referred to as the bond dimension of the MPO, denoted as $D_w = \max_l(\dim(b_l))$. Then $\Heff_c$ at the orthogonal center $c$ can be computed as
\begin{align}\label{eq:heff}
\Heff_{s_c, a_{c-1}, a_c}^{s_c', a_{c-1}', a_c'} = \sum_{b_{c-1}, b_c} L^{a_{c-1}'}_{b_{c-1}, a_{c-1}} O_{b_{c-1}, b_c}^{s_c, s_c'} R^{a_c'}_{b_c, a_c},
\end{align}
where $L$ and $R$ are rank-$3$ tensors which represent the effective environments left and right to the orthogonal center $c$. They can be computed iteratively
\begin{align}
& L^{a_l'}_{b_l, a_l} = \sum_{b_{l-1}, a_{l-1}, a_{l-1}', s_l, s_l'} L^{a_{l-1}'}_{b_{l-1}, a_{l-1}}  O_{b_{l-1}, b_l}^{s_l, s_l'} A_{a_{l-1}, a_l}^{s_l} (A_{a_{l-1}', a_l'}^{s_l'})^{\ast}; \label{eq:updateL} \\
& R^{a_{l-1}'}_{b_{l-1}, a_{l-1}} = \sum_{b_l, a_l, a_l', s_l, s_l'} R^{a_{l}'}_{b_{l}, a_{l}} O_{b_{l-1}, b_{l}}^{s_l, s_l'} B_{a_{l-1}, a_l}^{s_l} (B_{a_{l-1}', a_l'}^{s_l'})^{\ast}, \label{eq:updateR}
\end{align}
with the starting tensors
\begin{align}
&L^{a_1'}_{b_1, a_1} = \sum_{s_1, s_1'} O_{b_1}^{s_1, s_1'} A_{a_1}^{s_1} (A_{a_1'}^{s_1'})^{\ast}; \\
& R^{a_{L-1}'}_{b_{L-1}, a_{L-1}} = \sum_{s_L, s_L'} O_{b_{L-1}}^{s_L, s_L'} B_{a_{L-1}}^{s_L} (B_{a_{L-1}'}^{s_L'})^{\ast}.
\end{align}
Here we note that computing $L_{c-1}$ and $R_c$ for center $c$ requires to compute Eqs.(\ref{eq:updateL}, \ref{eq:updateR}) throughout the chain. However, computations can be saved if we first compute all the $R$ tensors before hand and store them in memory, then during the left to right sweep at site $c$, one only have to evaluate Eq.(\ref{eq:updateL}) for $l=c$ to update the storage. This can be done similarly during the right to left sweep. In this way one reduces the total number of evaluations of Eqs.(\ref{eq:updateL}, \ref{eq:updateR}) from $L^2$ to $2L$. Additionally, instead of building $\Heff$ explicitly as in Eq.(\ref{eq:heff}), one can simply implement its operation on an input rank-$3$ tensor $X$ with an output rank-$3$ tensor $Y$, that is $Y = \Heff(X)$, which is explicitly
\begin{align}\label{eq:heffop}
Y^{s_c'}_{a_{c-1}', a_c'} = \sum_{b_{c-1}, b_c, a_{c-1}, a_c, s_c} L^{a_{c-1}'}_{b_{c-1}, a_{c-1}} O_{b_{c-1}, b_c}^{s_c, s_c'} R^{a_c'}_{b_c, a_c} X^{s_c}_{a_{c-1},a_c}.
\end{align}
The complexity of evaluating Eq.(\ref{eq:heffop}) is $O(dD_wD^3)$. An iterative eigensolver is able to compute the lowest eigenpairs once the operation in Eq.(\ref{eq:heffop}) is given.

\section{Comparison between the low temperature results and the results from the ground state}
\begin{figure}
\includegraphics[width=\columnwidth]{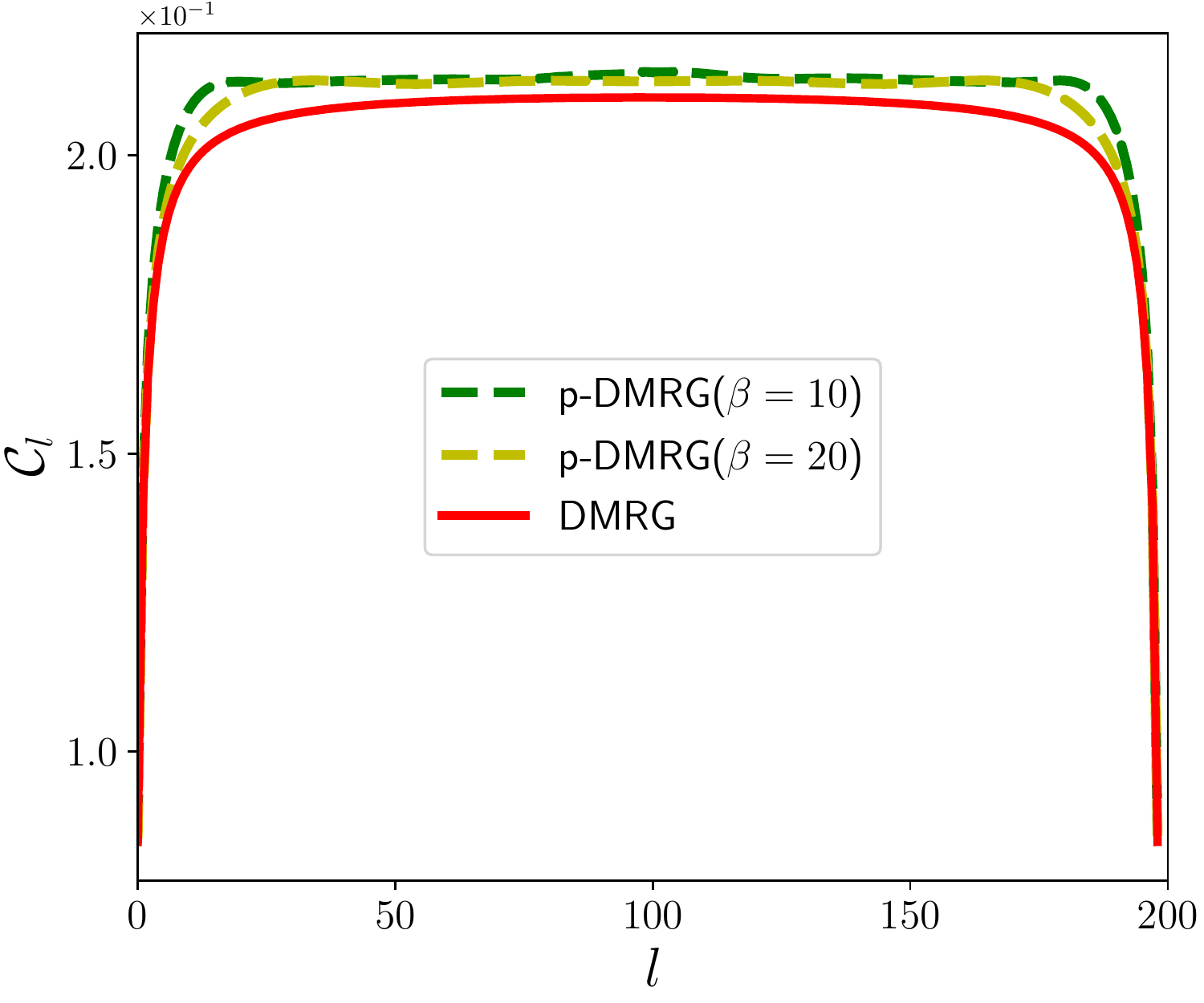}
\caption{Comparison between the low-temperature correlations and the ground state (zero-temperature) correlations for $L=200$. For both DMRG and p-DMRG we have used $D=30$. For p-DMRG, we have further used $R=50$ for $\beta=10$ and $R=10$ for $\beta=20$.
}
\label{fig:figS0}
\end{figure}

To demonstrate the effectiveness and efficiency of the p-DMRG algorithm for computing equilibrium states, we have used the transverse Ising chain as an example with the inverse temperatures $\beta=10,20$ in the main text. Here we also directly compare these low-temperature results to their ground state values (The ground state is computed by the standard DMRG), to show that they have a non-negligible derivation from the latter, which is shown in Fig.~\ref{fig:figS0}. We can see that the correlations corresponding to $\beta=10$ and $\beta=20$ are similar in the bulk and differs close to the boundaries, which they have a finite difference to the ground state values in both the bulk and the boundaries.

\section{Efficient construction of the local problem for computing the non-equilibrium steady states}
Now we assume that the Lindblad operator $\lindblad$ can be written as an MPO 
\begin{align}
\lindblad_{s_1,r_1,s_2,r_2, \dots, s_L,r_L}^{s_1',r_1',s_2',r_2', \dots, s_L',r_L'} =& \sum_{b_1, b_2, \dots, b_{L-1}} O_{b_1}^{s_1,r_1, s_1',r_1'} \nonumber \\ 
&O_{b_1, b_2}^{s_2,r_2, s_2',r_2'} \dots O_{b_{L-1}}^{s_L,r_L, s_L',r_L'}.
\end{align}
In the following we will use $x, y$ to denote the auxiliary indices of the positive matrix product ansatz.
Then $\Leff$ can be computed as
\begin{align}
\Leff_{s_c, r_c, x_{c-1}, y_{c-1}, x_c, y_c}^{s_c', r_c', x_{c-1}', y_{c-1}', x_c', y_c'} =& \sum_{b_{c-1}, b_c} L^{x_{c-1}', y_{c-1}'}_{b_{c-1}, x_{c-1}, y_{c-1}}  \nonumber \\ 
&O_{b_{c-1}, b_c}^{s_c,r_c, s_c',r_c'} R^{x_c', y_c'}_{b_{c}, x_c, y_c} .
\end{align}
Here we have used the same symbol $L$ and $R$, but they are not directly related to Eq.(\ref{eq:heff}). Similarly, the $L$ and $R$ tensors in this case can be iteratively computed as
\begin{align}
 L^{x_l', y_l'}_{b_l, x_l, y_l} =& \sum_{b_{l-1}, x_{l-1}, y_{l-1}, x_{l-1}', y_{l-1}', s_l,r_l, s_l',r_l'} \nonumber \\ 
 & L^{x_{l-1}', y_{l-1}'}_{b_{l-1}, x_{l-1}, y_{l-1}} O_{b_{l-1}, b_l}^{s_l,r_l, s_l', r_l'} A_{x_{l-1}, x_l}^{s_l} \nonumber \\ 
 & (A_{y_{l-1}, y_l}^{r_l})^{\ast} (A_{x_{l-1}', x_l'}^{s_l'})^{\ast} A_{y_{l-1}', y_l'}^{r_l'} ; \label{eq:updateL2} \\
 R^{x_{l-1}', y_{l-1}'}_{b_{l-1}, x_{l-1},y_{l-1}} =& \sum_{b_l, x_l, y_l, x_l',y_l', s_l,r_l, s_l',r_l'} R^{x_{l}', y_l'}_{b_{l}, x_{l}, y_l} O_{b_{l-1}, b_{l}}^{s_l,r_l, s_l',r_l'} \nonumber \\ 
 &B_{x_{l-1}, x_l}^{s_l} (B_{y_{l-1}, y_l}^{r_l})^{\ast} (B_{x_{l-1}', x_l'}^{s_l'})^{\ast} B_{y_{l-1}', y_l'}^{r_l'} , \label{eq:updateR2}
\end{align}
with the starting tensors
\begin{align}
L^{x_1',y_1'}_{b_1, x_1,y_1} =& \sum_{s_1,r_1, s_1',r_1'} O_{b_1}^{s_1,r_1, s_1',r_1'} A_{x_1}^{s_1} (A_{y_1}^{r_1})^{\ast} (A_{x_1'}^{s_1'})^{\ast} A_{y_1'}^{r_1'}; \\
 R^{x_{L-1}',y_{L-1}'}_{b_{L-1}, x_{L-1},y_{L-1}} =& \sum_{s_L,r_L, s_L',r_L'} O_{b_{L-1}}^{s_L,r_L, s_L',r_L'} B_{x_{L-1}}^{s_L} (B_{y_{L-1}}^{r_L})^{\ast} \nonumber \\ 
 & (B_{x_{L-1}'}^{s_L'})^{\ast} B_{y_{L-1}'}^{r_L'}.
\end{align}
Similar to the case of equilibrium states, one can first compute all the $R$ tensors and then update them one by one during each local optimization to reduce the total number of evaluations of Eqs.(\ref{eq:updateL2}, \ref{eq:updateR2}). For the local optimization, one should in general also treat $\Leff$ as a linear operation on an input rank-$6$ tensor $X$ with an output rank-$6$ tensor $Y$ ($Y = \Leff(X)$), which is explicitly
\begin{align}
Y^{s_c',r_c'}_{x_{c-1}',y_{c-1}', x_c',y_c'} =& \sum_{b_{c-1}, b_c, x_{c-1},y_{c-1},x_c,y_c, s_c, r_c} L^{x_{c-1}', y_{c-1}'}_{b_{c-1}, x_{c-1}, y_{c-1}} \nonumber \\ 
& O_{b_{c-1}, b_c}^{s_c,r_c, s_c',r_c'} R^{x_c',y_c'}_{b_c, x_c, y_c} X^{s_c,r_c}_{x_{c-1},y_{c-1},x_c,y_c}.
\end{align}

\section{More details of the numerical simulations of the boundary driven XXZ chain}\label{app:xxzmore}

\begin{figure}
\includegraphics[width=\columnwidth]{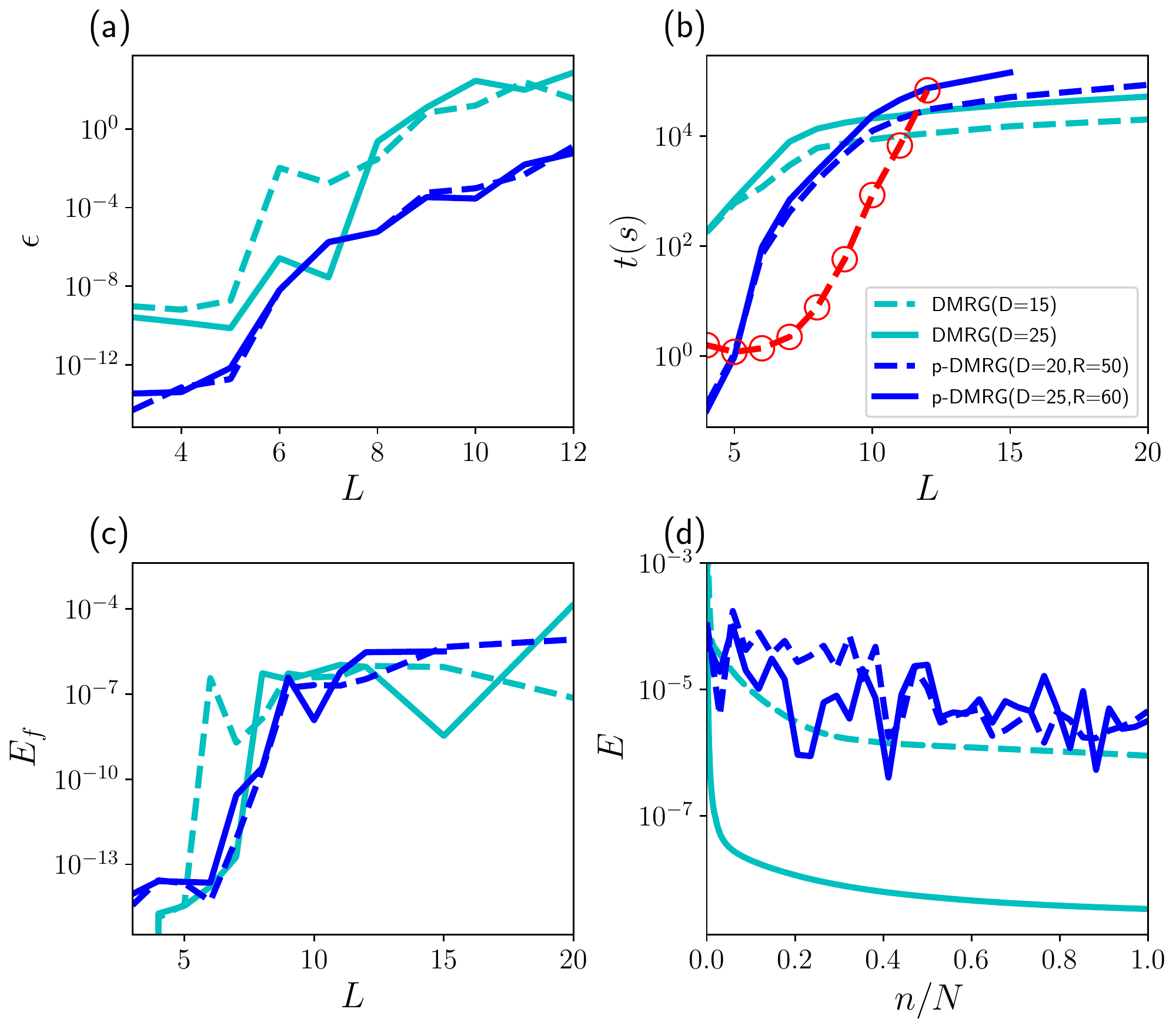}
\caption{(a) The mean relative error of the current, defined as $\epsilon = \sum_{l=1}^{L-1} |\current_l/\current - 1|/(L-1)$ with $\current$ the exact value from ED, as a function of the system size $L$. (b) The runtime scaling as a function of $L$. (c) The final energy as a function of $L$. (d) The energy versus the minimization step $n$ for $L=15$ ($N$ is the total number of minimization steps). In all panels the cyan dashed and solid lines represent DMRG results with $D=15$ and $D=25$ respectively, and the blue dashed and solid lines represent p-DMRG results with $D=20$ and $D=25$ respectively. For DMRG we have used $10000$ sweeps while for p-DMRG we have used $2$ sweeps. 
}
\label{fig:figS1}
\end{figure}

The boundary driven XXZ chain, due to the rapid vanishing of the spectrum gap for $\lindblad$ (typically $O(1/L^3)$), is extremely difficult to solve numerically. Nevertheless, this case can be analytically solved~\cite{Prosen2011b}. Therefore it is an ideal test ground for different numerical methods.

The additional details of our simulations for the boundary driven XXZ chain are shown in Fig.~\ref{fig:figS1}. In Fig.~\ref{fig:figS1}(a), we show the convergence of the DMRG and p-DMRG algorithm when increasing $D$ (and also increasing $R$ for p-DMRG). We can see that for both algorithms the relatively error $\epsilon$ compared to the exact values from ED does not improve significantly with $D$. For DMRG there is no clear improvement at $D=25$ and $\current$ does not converge as well starting from $L=8$ for $D=25$, same to the case of $D=15$. For p-DMRG we see tiny improvements starting from $L=8$ (except for the case with $L=11$ which may be due to a bad random initial state). In p-DMRG there is an additional source of error compared to the single-site DMRG, namely the SVD truncation used during center move (in single-site DMRG on truncation is done in the SVD used after local optimization). Since we are using a relatively small $D$ in both algorithms, we observe that the additional SVD truncation could easily induce an error of the order $10^{-5}$, which is the same order of $\current$ for large $L$. Thus a high-precision current can not be expected from p-DMRG if one does not increase $D$ significantly, which would be extremely expensive. Nevertheless, we can see that for this problem p-DMRG can already reach a much higher precision compared to DMRG in almost all the cases we have considered. For p-DMRG we can still obtain relatively reasonable current at $L=15$. In Fig.~\ref{fig:figS1}(b), we show the runtime scaling for DMRG and p-DMRG at different $D$s. First we see the exponential scaling of ED as expected. For p-DMRG the advantage compared to ED only appears from $L=12$ (but the scaling of DMRG or p-DMRG is of course more favorable if the issue of convergence is not considered). Here we note that for $\Delta=1$ when the transport is diffusive, it has been shown that one could compute the steady state current with fairly high precision for up to $100$ spins, using a specially designed initialization strategy~\cite{CasagrandeLandi2021}.

To better visualize the convergence for DMRG and p-DMRG, we further show the final energy $E_f$ as a function of $L$ in Fig.~\ref{fig:figS1}(c) and the energy $E$ as a function of the local minimization step for the particular case of $L=12$ in Fig.~\ref{fig:figS1}(d). From Fig.~\ref{fig:figS1}(c) we can see that for all the $2$ cases we have studied with p-DMRG, the final energy can only reach a value of the order $10^{-5}$, which is the reason that we could not expect to get a precise $\current$ for p-DMRG since one expects $\current$ to be the order $10^{-5}$ or less after $L=12$. For DMRG $\current$ is completely wrong even for $L=9,10$ where the final energies are of the order $10^{-7}$ (and for $L=15$ where the final energy is less than $10^{-8}$). The most important reason for this discrepancy between the energy and $\current$ could be that the resulting state is unphysical (which can be directly seen by checking with a Hermitian observable such as $\sgz_l$, the imaginary part of which will be significantly different from $0$). Another reason may be that the energy of $\lindblad^{\dagger}\lindblad$ loses the physical meaning of the spectrum of the original Lindblad operator $\lindblad$. From Fig.~\ref{fig:figS1}(d) we can see the extremely slow convergence of DMRG. Even the worse, although the energy could be significantly slower when we increase $D$ from $15$ to $25$ in DMRG, the predicted $\current$ is still completely wrong (even for the sign). For p-DMRG the results converge fairly well with only $2$ sweeps, although monotonic convergence is lost due to the effects explained in the main text. Larger $D$ and $R$ are required to reach lower energies for p-DMRG. Nevertheless, $\current$ at $L=15$ predicted with p-DMRG is still reasonable (although the error is not negligible).

In the end we note that due to the faster scaling of the complexity of p-DMRG compared to DMRG, the runtime for DMRG is not larger than p-DMRG even if the number of sweeps used for DMRG is much larger, which can be seen from Fig.~\ref{fig:figS1}(b). Actually the calculation becomes demanding when $L\geq 10$ for all the algorithms considered. The current issue with p-DMRG is that the local optimization is too expensive ($O(d^2D^6)$), partially due to that the low-Schmidt-rank nature of $\rhoeff$ is not made use of when solving the local problem, which we leave to future investigations.

\end{document}